\documentclass[10pt,aps,prb,twocolumn,groupedaddress,eqsecnum,draft,showpacs,intlimits,floatfix,amsmath,amssymb]{revtex4-1}
\usepackage{bm}
\usepackage[final]{graphicx}

\DeclareMathOperator{\Tr}{Tr}
\DeclareMathOperator{\Img}{Im}
\begin{document}

\title{Dynamical mean-field theory of correlated hopping: A rigorous local approach}

\author{Andrij M. Shvaika}

\affiliation{Institute for Condensed Matter Physics,
National Academy of Sciences of Ukraine, \\
1~Svientsitskii Str., UA--79011 Lviv, Ukraine}

\email{ashv@icmp.lviv.ua}

\homepage{http://ph.icmp.lviv.ua/~ashv/}

\date{\today}

\begin{abstract}
A general approach for the description of correlated hopping in infinite
dimensions, which is based on an expansion over electron hopping around
the atomic limit, is developed. Such an approach keeps the dynamical
mean-field theory local ideology and allows one to calculate the
thermodynamical functions. A grand-canonical potential functional and a
$\Phi$-derivatible theory that does not introduce the self-energy is
proposed. As limiting cases the Falicov--Kimball model with correlated
hopping and the model with broken bonds (``diluted" conductor) are
investigated, and the connection with the Blackman-Esterling-Berk coherent
potential approximation approach in the theory of the binary alloy with
off-diagonal disorder is considered.
\end{abstract}


\maketitle

\section{Introduction}\label{intro}

A microscopic origin of the different effects and phenomena specific to
transition and rare-earth metals and oxides, including high-$T_c$,
superconductors is due to the strong local correlations between electrons.
The simplest interaction that introduces such local correlations is the
Hubbard single-site Coulomb repulsion
$Un_{i\uparrow}n_{i\downarrow}$.\cite{H-I} In addition to such local terms
there are others involving neighboring lattice sites that describe the
charge-charge interaction $Vn_in_j$ and correlated or density-dependent
hopping:
\begin{equation}\label{ch1}
  \sum_{ij\sigma} t^{(1)}_{ij} a_{i\sigma}^{\dagger}a_{j\sigma}
  \left(n_{i\Bar\sigma}+n_{j\Bar\sigma}\right).
\end{equation}
In general, an effective model with correlated hopping can be obtained
after integrating out additional degrees of freedom\cite{Foglio,Simon1} or
as a phenomenological model. It was first proposed to describe the
properties of mixed valence systems, but after the discovery of high-$T_c$
superconductivity the interest in correlated hopping as a possible new
mechanism for a superconducting instability and temperature-induced
metal-insulator transition increased.\cite{Hirsch,Didukh,Simon1,Simon2} In
addition correlated hopping is considered an important factor in the
stabilization of the ferromagnetism and localization of
electrons.\cite{Kollar}

Models with correlated hopping, like other models with strong electron
correlations, cannot be solved exactly. However in the last decade, great
progress in the description of strongly correlated electron systems was
achieved with the development of the dynamical mean-field theory
(DMFT).\cite{MetznerVollhardt,DMFTreview}

The main idea of the dynamical mean-field theory, which is exact in the
limit of infinite dimensions, is in the local (single-site) nature of the
self-energy.\cite{DMFTreview} But this statement, as shown by
Schiller,\cite{Schiller} is violated for systems with correlated hopping,
when the self-energy becomes nonlocal that breaks the general DMFT scheme.
Now the effect of correlated hopping is similar to the Hartree
(mean-field) renormalization of the band energy, as an additional nonlocal
contribution to the self-energy.

A possible solution of this problem is in a reformulation of the DMFT
approach without introducing the self-energy. It should be noted that
self-energy appears in the Dyson equation for the one-electron Green's
function as a correction to the band energy due to many-electron
interactions, which are treated as perturbations. On the other hand, one
can start, not from one-electron band states, but from many-electron local
states, and analyze the influence of the intersite hopping $t_{ij}$ on the
creation of the band structure. Such a strong-coupling approach
corresponds to the perturbation theory over electron hopping. In this case
the one-electron Green's function can be a solution to the Larkin
equation,\cite{Larkin,Vaks}
\begin{equation}\label{LarkinIntro}
  \begin{split}
  G_{ij}(\omega)&=\Xi_{ij}(\omega)+\sum_{lm}\Xi_{il}(\omega)t_{lm}G_{mj}(\omega),\\
  G_{\bm{k}}(\omega)&=\Xi_{\bm{k}}(\omega)+\Xi_{\bm{k}}(\omega)t_{\bm{k}}G_{\bm{k}}(\omega),
  \end{split}
\end{equation}
where $\Xi_{ij}(\omega)$ is an irreducible part of the Green's function
that cannot be divided into parts by cutting one hopping line, and is
connected to the self-energy by the following equation:
\begin{equation}\label{XiSigma}
  \Xi_{\bm{k}}^{-1}(\omega)=\omega+\mu-\Sigma_{\bm{k}}(\omega).
\end{equation}
In the DMFT approach such an irreducible part (irreducible cumulant) was
intoduced by Metzner,\cite{Metzner} who proved that it is local in the
$D\to\infty$ limit. Below it will be shown that this statement is more
general than the one about the local nature of the self-energy.

In this paper we present a general approach for a description of
correlated hopping in infinite dimensions which is based on an expansion
over electron hopping around the atomic limit,\cite{Metzner,ShvaikaPRB}
which keeps the DMFT local ideology, and which allows one to calculate the
thermodynamical functions. As a limiting case the Falicov--Kimball model
with correlated hopping and a model with broken bonds (``diluted"
conductor) are considered. In this case, such a strong coupling approach
corresponds to the generalized locator coherent potential approximation
developed by Blackman, Esterling and Berk\cite{BEB} (BEB CPA) for the
description of the binary alloy with off-diagonal disorder.

\section{Expansion around the atomic limit}

In general, the hopping term of the Hamiltonian with correlated hopping
for the Falicov--Kimball and Hubbard models is written
\begin{equation}\label{HtFK}
 \begin{split}
  H_t=\frac{1}{\sqrt{D}}\sum_{\langle ij\rangle}\Bigl[ & t_1d_i^{\dag}d_j +
  t_2d_i^{\dag}d_j\left(n_{if}+n_{jf}\right)
  \\
  & + t_3d_i^{\dag}d_jn_{if}n_{jf}\Bigr]
 \end{split}
\end{equation}
and
\begin{equation}\label{HtH}
 \begin{split}
  H_t=\frac{1}{\sqrt{D}}\sum_{\substack{\langle ij\rangle\\ \sigma}} \Bigl[ & t_1a_{i\sigma}^{\dag}a_{j\sigma} +
  t_2a_{i\sigma}^{\dag}a_{j\sigma} \left(n_{i\Bar\sigma} + n_{j\Bar\sigma}\right)
  \\
  & + t_3a_{i\sigma}^{\dag}a_{j\sigma} n_{i\Bar\sigma}n_{j\Bar\sigma}
  \Bigr],
 \end{split}
\end{equation}
respectively. As a rule, the first two terms with $t_1$ and $t_2$ were
considered,\cite{Hirsch,Didukh,Kollar,Schiller} but an influence of the
last term with $t_3$, that can originate from the nondirect interactions
over other degrees of freedom, was also
investigated.\cite{Simon1,Simon2,Lemanski} In Hamiltonians (\ref{HtFK})
and (\ref{HtH}) the terms $t_1$, $t_2$, and $t_3$ describe hopping between
two nearest-neighbor sites $i$ and $j$ irrespective of the occupation of
these sites, the hopping between these sites when one of them is occupied
by another particle and the hopping between two occupied sites,
respectively. One can transfer Hamiltonian (\ref{HtFK}) to the other one
that describes the hopping between the sites with the certain occupations,
\begin{align}
  H_t&=\frac{1}{\sqrt{D}}\sum_{\langle ij\rangle}\Bigl[
  t_{ij}^{++}P_i^+d_i^{\dag}d_jP_j^+ + t_{ij}^{--}P_i^-d_i^{\dag}d_jP_j^-
  \nonumber\\
  &\qquad+t_{ij}^{+-}P_i^+d_i^{\dag}d_jP_j^- +
  t_{ij}^{-+}P_i^-d_i^{\dag}d_jP_j^+\Bigr]
  \label{HtFKG}\\
  &=\frac{1}{\sqrt{D}}\sum_{\langle ij\rangle}\mathbf{d}_i^{\dag}\mathbf{t}_{ij}\mathbf{d}_j,
  \nonumber
\end{align}
where the projection operators
\begin{equation}\label{P}
 P_i^+=n_{if},\quad    P_i^-=1-n_{if}
\end{equation}
and matrix notations
\begin{equation}\label{mtrx}
  \mathbf{d}_i=\begin{pmatrix} P_i^+ \\ P_i^- \end{pmatrix} d_i, \quad
  \mathbf{t}_{ij}=\begin{pmatrix}
                    t_{ij}^{++} & t_{ij}^{+-} \\
                    t_{ij}^{-+} & t_{ij}^{--}
                  \end{pmatrix}
\end{equation}
are introduced. For the Hubbard model it corresponds to the transition to
the Hubbard operators
\begin{equation}\label{HtHG}
  H_t=\frac{1}{\sqrt{D}}\sum_{\substack{\langle ij\rangle\\ \sigma}}
  \mathbf{a}_{i\sigma}^{\dag}\mathbf{t}_{ij}\mathbf{a}_{j\sigma},
  \quad \mathbf{a}_{i\sigma}=\begin{pmatrix} \sigma X_i^{\Bar\sigma2} \\
  X_i^{0\sigma} \end{pmatrix}.
\end{equation}

It is obvious that
\begin{equation}\label{t1}
  t_{ij}^{\alpha\gamma}=\left(t_{ji}^{\gamma\alpha}\right)^*
\end{equation}
and hopping integrals in Eqs. (\ref{HtFK}), (\ref{HtH}), and (\ref{mtrx})
are connected by the following equations
\begin{align}
  t^{--}&=t_1,             &t_1&=t^{--},
  \nonumber\\
  t^{+-}&=t^{-+}=t_1+t_2,  &t_2&=t^{+-(-+)}-t^{--},
  \label{t2}\\
  t^{++}&=t_1+2t_2+t_3,    &t_3&=t^{++}+t^{--}-t^{+-}-t^{-+}.
  \nonumber
\end{align}

The total Hamiltonian of the electronic system with local correlations and
correlated hopping,
\begin{equation}\label{Htot}
  H=\sum_i H_i + H_t,
\end{equation}
includes, besides the hopping term $H_t$ that is not local, the
single-site contributions $H_i$. Our aim is to consider the $D\to\infty$
limit, and, as mentioned in Sec.~\ref{intro}, it is convenient to start
not from the Dyson equation, that considers the terms with correlated
hopping $t_2$ and $t_3$ as some kind of many-particle interactions, but
from the Larkin equation\cite{Larkin,Vaks} in a coordinate representation,
\begin{equation}\label{Larkin}
  \mathbf{G}_{ij}(\omega)=\mathbf{\Xi}_{ij}(\omega)+
  \sum_{lm}\mathbf{\Xi}_{il}(\omega)\mathbf{t}_{lm}\mathbf{G}_{mj}(\omega),
\end{equation}
that treats all hopping terms in the same manner. Here an irreducible part
(irreducible cumulant\cite{Metzner}) $\mathbf{\Xi}_{ij}(\omega)$, that
cannot be divided into parts by cutting one hopping line
$\mathbf{t}_{lm}$, is introduced.

For the models with correlated hopping all quantities in Eq.
(\ref{Larkin}) are matrices, e.g., the components of the Green's function
$\mathbf{G}_{ij}(\omega)$ are constructed by the projected (Hubbard)
operators
\begin{equation}\label{GFcomp}
  \begin{split}
  G_{ij}^{\alpha\gamma}(\tau-\tau')&=-\left\langle T (P_i^{\alpha} d_i)_{\tau}
  (d_j^{\dag}P_j^{\gamma})_{\tau'}\right\rangle=
  \beta\frac{\delta\Omega}{\delta t_{ji}^{\gamma\alpha}(\tau'-\tau)},
  \\
  \mathbf{G}_{ij}(\tau-\tau')&=-\left\langle T \mathbf{d}_i(\tau)\otimes
  \mathbf{d}_j^{\dag}(\tau')\right\rangle=
  \beta\frac{\delta\Omega}{\delta \mathbf{t}_{ji}(\tau'-\tau)},
  \end{split}
\end{equation}
where $\Omega$ is a grand canonical potential functional and $\delta
t_{ji}^{\gamma\alpha}(\tau'-\tau)$ is some auxiliary field.

After the Fourier transformation to the momentum space one can find the
Green's function as a solution of Larkin equation (\ref{Larkin}),
\begin{equation}\label{Larkinsol}
\begin{split}
  \mathbf{G}_{\bm{k}}(\omega)&=
  \left[\openone - \mathbf{\Xi}_{\bm{k}}(\omega) \mathbf{t}_{\bm{k}}\right]^{-1}
  \mathbf{\Xi}_{\bm{k}}(\omega)\\
  &=\left[\mathbf{\Xi}_{\bm{k}}^{-1}(\omega) - \mathbf{t}_{\bm{k}}\right]^{-1},
\end{split}
\end{equation}
and the total physical Green's function
\begin{equation}
  G_{ij}(\tau-\tau')=-\left\langle T d_i(\tau) d_j^{\dag}(\tau')\right\rangle
  =\frac{\delta\Omega}{\delta t_{1\,ji}(\tau'-\tau)}
\end{equation}
is equal to the sum of all matrix components
\begin{equation}\label{GFtot}
  G_{ij}(\omega)=\sum_{\alpha\gamma} G_{ij}^{\alpha\gamma}(\omega).
\end{equation}

In general, an irreducible part $\mathbf{\Xi}_{ij}(\omega)$ is represented
diagrammatically by the single-site vertices (because all interactions in
$H_i$ are local) connected by hopping lines\cite{ShvaikaPRB} and in the
$D\to\infty$ limit it can be shown (see, e.g., Ref.~\onlinecite{Metzner})
that it becomes local
\begin{equation}\label{Xiloc}
  \mathbf{\Xi}_{ij}(\omega)=\delta_{ij}\mathbf{\Xi}(\omega), \quad
  \mathbf{\Xi}_{\bm{k}}(\omega)=\mathbf{\Xi}(\omega).
\end{equation}

Such a matrix representation allows one to reformulate the dynamical
mean-field theory of systems with correlated hopping in terms of local
quantities. Indeed, the local irreducible part $\mathbf{\Xi}(\omega)$
depends on the electron hopping only via the local coherent potential
(auxiliary $\lambda$ field of Brandt and Mielsch\cite{Brandt})
\begin{equation}\label{CohPotcomp}
  J^{\alpha\epsilon}(\omega)=\sum_{lm\gamma\delta}t_{ol}^{\alpha\gamma}\,
  G_{lm}^{[o]\gamma\delta}(\omega)\, t_{mo}^{\delta\epsilon}
\end{equation}
or, in matrix notations,
\begin{equation}\label{CohPot}
  \mathbf{J}(\omega)=\sum_{lm}\mathbf{t}_{ol}\,
  \mathbf{G}_{lm}^{[o]}(\omega)\, \mathbf{t}_{mo},
\end{equation}
where $\mathbf{G}_{lm}^{[o]}(\omega)$ is the Green's function for lattice
with the removed site $o$. On the other hand, we can introduce another
quantity
\begin{equation}\label{I}
  \mathbf{I}(\omega)=\sum_{lm}\mathbf{t}_{ol}\,
  \mathbf{G}_{lm}(\omega)\, \mathbf{t}_{mo}=
  \frac1N\sum_{\bm{k}}\mathbf{t}_{\bm{k}}\,
  \mathbf{G}_{\bm{k}}(\omega)\, \mathbf{t}_{\bm{k}}
\end{equation}
that is connected with coherent potential [Eq. (\ref{CohPot})] by the
equation
\begin{equation}\label{IvsJ}
  \mathbf{I}(\omega)=\left[\openone-\mathbf{J}(\omega)\mathbf{\Xi}(\omega)\right]^{-1}
  \mathbf{J}(\omega).
\end{equation}
From Eqs. (\ref{CohPot}), (\ref{I}) and (\ref{IvsJ}) we obtain the
following equation for the coherent potential $\mathbf{J}(\omega)$:
\begin{equation}\label{DMFTeq1}
  \frac1N\sum_{\bm{k}}
  \left[\mathbf{\Xi}^{-1}(\omega) - \mathbf{t}_{\bm{k}}\right]^{-1}
  =\left[\mathbf{\Xi}^{-1}(\omega) - \mathbf{J}(\omega)\right]^{-1},
\end{equation}
that is a matrix generalization of the known one (see, e.g.,
Ref.~\onlinecite{DMFTreview}).

The expression in the right-hand side of Eq. (\ref{DMFTeq1}) is the Larkin
representation of the Green's function,
\begin{equation}\label{DMFTeq2}
  \mathbf{G}_{\text{imp}}(\omega)
  =\left[\mathbf{\Xi}^{-1}(\omega) - \mathbf{J}(\omega)\right]^{-1},
\end{equation}
for the effective single-impurity problem with the statistical operator
\begin{equation}\label{rhoimp}
  \Hat{\rho}_{\text{imp}}=e^{-\beta H_o}T\exp\left\{\!
  -\!\int_0^{\beta}\!\!d\tau\!\int_0^{\beta}\!\!d\tau'
  \mathbf{d}_o^{\dag}(\tau)\mathbf{J}(\tau-\tau')\mathbf{d}_o(\tau')
  \!\right\}.
\end{equation}
Equations (\ref{DMFTeq1}) and (\ref{DMFTeq2}) give the closed set of
equations for the \emph{local} coherent potential $\mathbf{J}(\omega)$ and
irreducible part $\mathbf{\Xi}(\omega)$,
\begin{gather}
  \frac1N\sum_{\bm{k}}
  \left[\mathbf{G}_{\text{imp}}^{-1}(\omega) + \mathbf{J}(\omega)
  - \mathbf{t}_{\bm{k}}\right]^{-1}=
  \mathbf{G}_{\text{imp}}(\omega),
  \label{DMFTeqJ}\\
  \mathbf{\Xi}^{-1}(\omega)=
  \mathbf{G}_{\text{imp}}^{-1}(\omega) + \mathbf{J}(\omega),
  \label{Xigensol}
\end{gather}
that was derived without introducing of the self-energy for the total
Green's function [Eq. (\ref{GFtot})]. Solution (\ref{Xigensol}) is
substituted into the expression for the lattice Green's function [Eq.
(\ref{Larkinsol})] that allows one to calculate the total Green's function
(\ref{GFtot}). Below it will be shown for the Falicov--Kimball model with
correlated hopping how to solve the effective single-impurity problem, to
calculate $\mathbf{G}_{\text{imp}}(\omega)$, and to derive the self-energy
that is now momentum dependent.

Another advantage of such an approach is in the correct analytic
properties of all quantities. In the complex plane they behave as
\begin{align}
  \mathbf{G}(z) & \to\frac{\mathbf{C}_{G}}{z}, \nonumber\\
  \mathbf{J}(z) & \to\frac{\mathbf{C}_{J}}{z}, \label{omginf}\\
  \mathbf{\Xi}(z) & \to\frac{\mathbf{C}_{\Xi}}{z} \nonumber
\end{align}
for $|z|\to\infty$, that allows one to build the Lehmann representation
for the Green's functions
\begin{gather}\label{Gimplem}
  \mathbf{G}_{\text{imp}}(z)=\frac1{\pi}\int_{-\infty}^{+\infty}\!d\omega
  \frac{\Img\mathbf{G}_{\text{imp}}(\omega-i0^+)}{z-\omega},
  \\
  \mathbf{G}_{\bm{k}}(z)=\frac1{\pi}\int_{-\infty}^{+\infty}\!d\omega
  \frac{\Img\mathbf{G}_{\bm{k}}(\omega-i0^+)}{z-\omega},
  \label{Gklem}
\end{gather}
the irreducible part
\begin{equation}\label{Xilem}
  \mathbf{\Xi}(z)=\frac1{\pi}\int_{-\infty}^{+\infty}\!d\omega
  \frac{\Img\mathbf{\Xi}(\omega-i0^+)}{z-\omega},
\end{equation}
and the coherent potential
\begin{equation}\label{Jlem}
  \mathbf{J}(z)=\frac1{\pi}\int_{-\infty}^{+\infty}\!d\omega
  \frac{\Img\mathbf{J}(\omega-i0^+)}{z-\omega}.
\end{equation}
The last two equations follows from the definitions of the irreducible
part [Eq. (\ref{Larkin})] and the coherent potential [Eq. (\ref{CohPot})].
It should be noted that we can, if we wish, introduce a matrix of the
local (CPA) self-energies\cite{BEB} in the same way as in Eq.
(\ref{XiSigma}) by
\begin{equation}\label{locSigm}
  \mathbf{\Sigma}(z)=(z+\mu)\openone-\mathbf{\Xi}^{-1}(z),
\end{equation}
but components of such self-energy matrix do not possess correct analytic
properties because $C_{\Xi}^{\alpha\alpha}\ne1$ and they diverge for
$|z|\to\infty$ (see below for the Falicov--Kimball model).

\section{Summation over wave vector}

In Eq. (\ref{DMFTeq1}) the sum over wave vector $\bm{k}$ can be calculated
directly when we put
\begin{equation}\label{tvsa}
  \mathbf{t}_{ij}=\mathbf{a}t_{ij},
\end{equation}
where $\mathbf{a}$ is a matrix of the correlated hopping constants. In
this case, in (\ref{DMFTeq1}), we can replace the sum over $\bm{k}$ by the
integral with some density of states $\rho(t)$
\begin{align}
  \frac1N\sum_{\bm{k}}&
  \left[\mathbf{\Xi}^{-1}(\omega) - \mathbf{a}t_{\bm{k}}\right]^{-1}
  =
  \int_{-\infty}^{+\infty}\!dt\rho(t)
  \left[\mathbf{\Xi}^{-1}(\omega) - \mathbf{a}t\right]^{-1}
  \nonumber\\
  &=
  \int_{-\infty}^{+\infty}\!dt\rho(t)
  \left[\mathbf{a}^{-1}\mathbf{\Xi}^{-1}(\omega) -
  t\openone\right]^{-1}\mathbf{a}^{-1}.
  \label{dos}
\end{align}
Matrix $\mathbf{M}=\mathbf{a}^{-1}\mathbf{\Xi}^{-1}(\omega)$ can be
diagonalized by some transformation $\mathbf{V}$,
\begin{equation}\label{unit}
  \mathbf{V}^{-1}\mathbf{M}\mathbf{V}=
  \begin{Vmatrix}
  z_1&&0\\
  &\ddots&\\
  0&&z_s
  \end{Vmatrix},
\end{equation}
and finally we obtain
\begin{gather}
  \mathbf{G}_{\text{imp}}\mathbf{a}=\mathbf{V}
  \begin{Vmatrix}
  F(z_1)&&0\\
  &\ddots&\\
  0&&F(z_s)
  \end{Vmatrix}
  \mathbf{V}^{-1},
  \label{Gsumk}\\
  \mathbf{J}=\mathbf{a}\mathbf{V}
  \begin{Vmatrix}
  z_1-\frac1{F(z_1)}&&0\\
  &\ddots&\\
  0&&z_s-\frac1{F(z_s)}
  \end{Vmatrix}
  \mathbf{V}^{-1},
  \label{Jsumk}
\end{gather}
where
\begin{equation}\label{Hilb}
  F(z)=\int_{-\infty}^{+\infty}\! dt\frac{\rho(t)}{z-t}
\end{equation}
is the Hilbert transform of the density of states $\rho(t)$.

Now, let us consider some popular densities of states. The Hilbert
transform of the Lorentzian density of states
\begin{equation}\label{Lorentz}
  \rho(t)=\frac{W}{\pi}\frac1{W^2+t^2}
\end{equation}
is equal to
\begin{equation}\label{HLorentz}
  F(z)=\frac1{z\mp iW},
\end{equation}
that immediately gives
\begin{equation}\label{Jlorntz}
  \mathbf{J}(\omega)=\pm iW\mathbf{a}.
\end{equation}

For a Bethe lattice with a semielliptic density of
states,\cite{DMFTreview}
\begin{equation}\label{Bethe}
  \rho(t)=\frac2{\pi W^2}\sqrt{W^2-t^2},
\end{equation}
we have
\begin{gather}
  F(z)=\frac2{W^2}\left(z-\sqrt{z^2-W^2}\right),\\
  z-\frac1{F(z)}=\frac{W^2}4 F(z)
\end{gather}
that gives a matrix generalization of the known
re\-la\-tion\cite{Kajueter}
\begin{equation}\label{Jbethe}
  \mathbf{J}(\omega)=\frac{W^2}4\mathbf{a}\mathbf{G}_{\text{imp}}(\omega)\mathbf{a}.
\end{equation}

It should be noted that Eqs. (\ref{Gsumk}) and (\ref{Jsumk}) are correct
only when $\det\mathbf{a}\ne0$. In other case, a direct decomposition of
Eq. (\ref{dos}) into simple fractions over $t_{\bm{k}}$ must be used. But
results for the particular cases [Eqs. (\ref{Jlorntz}) and (\ref{Jbethe})]
are always correct.

For a density of states different than the Lorentzian and semielliptic
ones the iterative procedure is as follows. We start from the initial
value for the coherent potential $\mathbf{J}(\omega)$, e.g., that of the
Lorentzian density of states [Eq. (\ref{Jlorntz})], that is used to
calculate the Green's function $\mathbf{G}_{\text{imp}}(\omega)$ for the
single impurity problem [Eq. (\ref{rhoimp})] (see below for the
Falicov-Kimball model). Then Eq. (\ref{Xigensol}) is used to find an
irreducible part $\mathbf{\Xi}(\omega)$, Eq. (\ref{DMFTeqJ}) is employed
to find a new value of $\mathbf{G}_{\text{imp}}(\omega)$, and then Eq.
(\ref{Xigensol}) is used to find a new coherent potential
$\mathbf{J}(\omega)$ value. The algorithm is then repeated until its
converges.

\section{Thermodynamics and $\Phi$-derivatible theory}

One can see that the DMFT system of equations presented above is a matrix
generalization of the known equations for the systems without correlated
hopping, with the replacement of the local self-energy by the local
irreducible part
\begin{equation}\label{SEtoXi}
  \Sigma(\omega)=\omega+\mu-\Xi^{-1}(\omega).
\end{equation}
However, there are principal difficulties in obtaining expression for the
grand canonical potential (free energy).

The main idea in derivation of a grand canonical potential in the DMFT is
the following.\cite{Brandt,DMFTreview} We start from the Baym--Kadanoff
functionals for the lattice,
\begin{equation}\label{BKlat}
\begin{split}
  \frac{\Omega_{\text{lat}}}{N}=&-\frac1{\beta}\sum_{\nu}\frac1N\sum_{\bm{k}}
  \ln\left[i\omega_{\nu}+\mu-\Sigma(i\omega_{\nu})-t_{\bm{k}}\right]\\
  &-\frac1{\beta}\sum_{\nu}\frac1N\sum_{\bm{k}}
  G_{\bm{k}}(i\omega_{\nu})\Sigma(i\omega_{\nu})+\frac{\Phi_{\text{lat}}}N,
 \end{split}
\end{equation}
and the effective single-impurity problem:
\begin{equation}\label{BKimp}
\begin{split}
  \Omega_{\text{imp}}=&-\frac1{\beta}\sum_{\nu}
  \ln\left[i\omega_{\nu}+\mu-\Sigma(i\omega_{\nu})-J(i\omega_{\nu})\right]\\
  &-\frac1{\beta}\sum_{\nu}
  G_{\text{imp}}(i\omega_{\nu})\Sigma(i\omega_{\nu})+\Phi_{\text{imp}}.
 \end{split}
\end{equation}
In the case of the absence of correlated hopping the self-energies for the
lattice and impurity are local and the same. In addition
\begin{equation}\label{PhiPhi}
  \frac{\Phi_{\text{lat}}}N=\Phi_{\text{imp}}
\end{equation}
and
\begin{equation}\label{GG}
  \frac1N\sum_{\bm{k}}G_{\bm{k}}(i\omega_{\nu})=G_{\text{imp}}(i\omega_{\nu}),
\end{equation}
that allows one to exclude an unknown Luttinger--Ward functional and to
find the final expression
\begin{align}
  \frac{\Omega_{\text{lat}}}{N}=\Omega_{\text{imp}}
  -\frac1{\beta}\sum_{\nu}
  \biggl\{&\frac1N\sum_{\bm{k}}
  \ln\left[i\omega_{\nu}+\mu-\Sigma(i\omega_{\nu})-t_{\bm{k}}\right]
  \nonumber\\
  &+\ln G_{\text{imp}}(i\omega_{\nu})
  \biggr\}.
  \label{BKfin}
\end{align}
But in the case of correlated hopping the self-energy is momentum
dependent\cite{Schiller} and it is impossible to write down the
Baym--Kadanoff functional for the impurity in the form of Eq.
(\ref{BKimp}), and that is the main problem.

In order to solve this problem let us construct the Baym--Kadanoff-type
functional without introducing the self-energy. To do this we start from
the expression for the Green's function (\ref{Larkinsol}) and
(\ref{GFcomp}) in the Larkin representation
\begin{equation}
  \left\lVert G_{\bm{k}}^{\alpha\gamma}(\omega)\right\rVert=
  \left\lVert
  \beta\frac{\delta\Omega_{\text{lat}}}{\delta t_{\bm{k}}^{\gamma\alpha}(\omega)}
  \right\rVert=
  \left[\openone - \mathbf{\Xi}_{\bm{k}}(\omega) \mathbf{t}_{\bm{k}}\right]^{-1}
  \mathbf{\Xi}_{\bm{k}}(\omega),
\end{equation}
where we introduce an auxiliary field $\mathbf{t}_{\bm{k}}(\omega) =
\mathbf{t}_{\bm{k}} + \delta\mathbf{t}_{\bm{k}}(\omega)$, and a known
relation
\begin{equation}\label{algebr}
  \delta\ln\det\mathbf{A}=\Tr\left(\mathbf{A}^{-1}\delta\mathbf{A}\right).
\end{equation}
That gives
\begin{align}
  \frac{\Omega_{\text{lat}}}{N}=&\Omega_0+\frac{\Omega'}N\\
  &-\frac1{\beta}\sum_{\nu}\frac1N\sum_{\bm{k}}
  \ln\det\left[\openone-\mathbf{\Xi}_{\bm{k}}(i\omega_{\nu})
  \mathbf{t}_{\bm{k}}(i\omega_{\nu})\right] ,
  \nonumber
\end{align}
where $\Omega_0$ is the grand canonical potential for the single-site
Hamiltonian $H_i$ ($\mathbf{t}_{ij}=0$) and
\begin{equation}
 \begin{split}
  \beta\frac{\delta\Omega'}{\delta
  t_{\bm{k}}^{\gamma\alpha}(i\omega_{\nu})}&=
  -\sum_{\nu'\bm{k}'}
  \Tr\biggl\{
  \frac{\delta\mathbf{\Xi}_{\bm{k}'}(i\omega_{\nu'})}
  {\delta t_{\bm{k}}^{\gamma\alpha}(i\omega_{\nu})}
  \\
  \times\mathbf{t}_{\bm{k}'}&(i\omega_{\nu'})
  \left[\openone-\mathbf{\Xi}_{\bm{k}'}(i\omega_{\nu'})
  \mathbf{t}_{\bm{k}'}(i\omega_{\nu'})\right]^{-1}
  \biggr\}.
 \end{split}
\end{equation}
As mentioned above, the irreducible part
$\mathbf{\Xi}_{\bm{k}}(i\omega_{\nu})$ can be diagrammatically represented
by the single-site vertices connected by hopping lines
$\mathbf{t}_{\bm{k}}(i\omega_{\nu})$.\cite{ShvaikaPRB} On the other hand,
by summing up the series of the diagrams of the same topology, one can see
that hopping lines are collected into chains,
\begin{equation}\label{tchains}
 \begin{split}
  \Tilde{\mathbf{t}}_{\bm{k}}(i\omega_{\nu})&=
  \mathbf{t}_{\bm{k}}(i\omega_{\nu})+
  \mathbf{t}_{\bm{k}}(i\omega_{\nu})
  \mathbf{\Xi}_{\bm{k}}(i\omega_{\nu})
  \mathbf{t}_{\bm{k}}(i\omega_{\nu})+\cdots\\
  &=\mathbf{t}_{\bm{k}}(i\omega_{\nu})
  \left[\openone-\mathbf{\Xi}_{\bm{k}}(i\omega_{\nu})
  \mathbf{t}_{\bm{k}}(i\omega_{\nu})\right]^{-1},
 \end{split}
\end{equation}
and the irreducible part can be represented by some skeletal diagrams with
the $\mathbf{t}_{\bm{k}}(i\omega_{\nu})$ replaced by
$\Tilde{\mathbf{t}}_{\bm{k}}(i\omega_{\nu})$ (spreading of lines). Now we
can represent $\Omega'$ as
\begin{equation}
  \frac{\Omega'}N=-\frac1{\beta}\sum_{\nu}\frac1N\sum_{\bm{k}}
  \Tr\left[\mathbf{\Xi}_{\bm{k}}(i\omega_{\nu})
  \Tilde{\mathbf{t}}_{\bm{k}}(i\omega_{\nu})
  \right]+\frac{\Tilde{\Phi}_{\text{lat}}}N,
\end{equation}
where $\Tilde{\Phi}_{\text{lat}}$ is a Luttinger--Ward-type functional:
\begin{equation}\label{LWlat}
  \beta\frac{\delta\Tilde{\Phi}_{\text{lat}}}{\delta
  \Tilde{t}_{\bm{k}}^{\gamma\alpha}(i\omega_{\nu})}=
  \Xi_{\bm{k}}^{\alpha\gamma}(i\omega_{\nu}).
\end{equation}

Finally, for the grand canonical potential functional for the lattice we
obtain
\begin{equation}\label{BKSfin}
 \begin{split}
  \frac{\Omega_{\text{lat}}}{N}=&\Omega_0-\frac1{\beta}\sum_{\nu}\frac1N\sum_{\bm{k}}
  \ln\det\left[\openone-\mathbf{\Xi}_{\bm{k}}(i\omega_{\nu})
  \mathbf{t}_{\bm{k}}(i\omega_{\nu})\right]\\
  &-\frac1{\beta}\sum_{\nu}\frac1N\sum_{\bm{k}}
  \Tr\left[\mathbf{\Xi}_{\bm{k}}(i\omega_{\nu})
  \Tilde{\mathbf{t}}_{\bm{k}}(i\omega_{\nu})
  \right]+\frac{\Tilde{\Phi}_{\text{lat}}}N.
 \end{split}
\end{equation}

In the limit of infinite dimensions an irreducible part becomes local [Eq.
(\ref{Xiloc})], and from Eq. (\ref{DMFTeq1}) it follows that
\begin{equation}\label{lvst}
  \frac1N\sum_{\bm{k}}\Tilde{\mathbf{t}}_{\bm{k}}(i\omega_{\nu})=
  \Tilde{\mathbf{J}}(i\omega_{\nu}),
\end{equation}
where
\begin{equation}
  \Tilde{\mathbf{J}}(i\omega_{\nu})=
  \mathbf{J}(i\omega_{\nu})
  \left[\openone-\mathbf{\Xi}(i\omega_{\nu})
  \mathbf{J}(i\omega_{\nu})\right]^{-1}.
\end{equation}
On the other hand, we can write a functional for the impurity in the same
form,
\begin{equation}
 \begin{split}
  \Omega_{\text{imp}}=&\Omega_0-\frac1{\beta}\sum_{\nu}
  \ln\det\left[\openone-\mathbf{\Xi}(i\omega_{\nu})
  \mathbf{J}(i\omega_{\nu})\right]\\
  &-\frac1{\beta}\sum_{\nu}
  \Tr\left[\mathbf{\Xi}(i\omega_{\nu})
  \Tilde{\mathbf{J}}(i\omega_{\nu})
  \right]+\Tilde{\Phi}_{\text{imp}},
 \end{split}
\end{equation}
where
\begin{equation}\label{LWimp}
  \beta\frac{\delta\Tilde{\Phi}_{\text{imp}}}{\delta
  \Tilde{J}^{\gamma\alpha}(i\omega_{\nu})}=
  \Xi^{\alpha\gamma}(i\omega_{\nu}).
\end{equation}
From Eqs. (\ref{Xiloc}), (\ref{LWlat}), (\ref{LWimp}), and (\ref{lvst}) it
is easy to prove that
\begin{equation}
  \frac{\Tilde{\Phi}_{\text{lat}}}N=\Tilde{\Phi}_{\text{imp}},
\end{equation}
and finally we obtain an expression for the grand canonical potential for
the lattice in terms of the quantities for the impurity problem
\begin{equation}\label{latvsimp}
 \begin{split}
  \frac{\Omega_{\text{lat}}}{N}=
  \Omega_{\text{imp}}-\frac1{\beta}\sum_{\nu}
  \biggl\{&\frac1N\sum_{\bm{k}}
  \ln\det\left[\openone-\mathbf{\Xi}(i\omega_{\nu})
  \mathbf{t}_{\bm{k}}\right]\\
  &-\ln\det\left[\openone-\mathbf{\Xi}(i\omega_{\nu})
  \mathbf{J}(i\omega_{\nu})\right]
  \biggr\} .
 \end{split}
\end{equation}
In the absence of correlated hopping this reduces to the known expression
[Eq. (\ref{BKfin})].

Logarithms in Eq. (\ref{latvsimp}) originate from the sum of the loop
diagrams,
\begin{equation}
  \Tr\sum_l\frac1l\left[\mathbf{\Xi}(i\omega_{\nu})\mathbf{t}_{\bm{k}}\right]^l
\end{equation}
and
\begin{equation}
  \Tr\sum_l\frac1l\left[\mathbf{\Xi}(i\omega_{\nu})\mathbf{J}(i\omega_{\nu})\right]^l,
\end{equation}
and using analytic properties of the irreducible part [Eq. (\ref{Xilem})]
and the coherent potential [Eq. (\ref{Jlem})], in Eq. (\ref{latvsimp}) one
can replace the sum over Matsubara's frequencies by an integral over real
axis, that gives
\begin{equation}\label{latvsimpreal}
 \begin{split}
  \frac{\Omega_{\text{lat}}}{N}&=
  \Omega_{\text{imp}}-\frac1{\pi}\int_{-\infty}^{+\infty}\!
  \frac{d\omega}{e^{\beta\omega}+1}\\
  &\times\biggl\{\frac1N\sum_{\bm{k}}\Img
  \ln\det\left[\openone-\mathbf{\Xi}(\omega-i0^+)
  \mathbf{t}_{\bm{k}}\right]\\
  &-\Img\ln\det\left[\openone-\mathbf{\Xi}(\omega-i0^+)
  \mathbf{J}(\omega-i0^+)\right]
  \biggr\} .
 \end{split}
\end{equation}

Expressions (\ref{BKSfin}) and (\ref{LWlat}) are very similar to the one
known in the Baym--Kadanoff approach\cite{Baym} but now, instead of the
conjugated quantities the Green's function $G_{\bm{k}}(i\omega_{\nu})$ and
self-energy $\Sigma_{\bm{k}}(i\omega_{\nu})$, we have the renormalized
hopping $\Tilde{\mathbf{t}}_{\bm{k}}(i\omega_{\nu})$ and the irreducible
part $\mathbf{\Xi}_{\bm{k}}(i\omega_{\nu})$. Such a formal analogy allows
us to build a $\Phi$-derivatible theory starting from functional
(\ref{BKSfin}) in the same way as done by Baym and Kadanoff.

First of all, for the electron concentration we have
\begin{equation}\label{n}
  \begin{split}
  n_1&=-\frac1N\frac{d\Omega_{\text{lat}}}{d\mu_1}
  =\left.-\frac{d\Omega_{\text{imp}}}{d\mu_1}\right|_{\mathbf{J}=\mathrm{const}}\\
  &=-\frac{d\Omega_0}{d\mu_1}-\frac1N\frac{\partial\Tilde{\Phi}_{\text{lat}}}{\partial\mu_1},
  \end{split}
\end{equation}
where a partial derivative is taken over $\mu$ not in chains
$\Tilde{\mathbf{t}}_{\bm{k}}(i\omega_{\nu})$ [Eq. (\ref{tchains})]:
\begin{equation}\label{partial}
  \frac{d}{d\mu}=\frac{\partial}{\partial\mu}+
  \sum_{\alpha\gamma\nu\bm{k}}\frac{d\Tilde{t}_{\bm{k}}^{\alpha\gamma}(i\omega_{\nu})}{d\mu}
  \frac{\delta}{\delta\Tilde{t}_{\bm{k}}^{\alpha\gamma}(i\omega_{\nu})}.
\end{equation}

Next, for the correlation function (dynamical susceptibility)
\begin{equation}\label{Suscdef}
  \chi_{12}=\frac{dn_1}{d\mu_2}=\frac{dn_2}{d\mu_1}=
  -\frac1N\frac{d^2\Omega_{\text{lat}}}{d\mu_1 d\mu_2}
\end{equation}
we obtain an expression that can be represented diagrammatically
as\cite{ShvaikaJPS}
\begin{equation}\label{L-suscept}
 \chi_{12}= \raisebox{-9pt}{\includegraphics[scale=0.3]{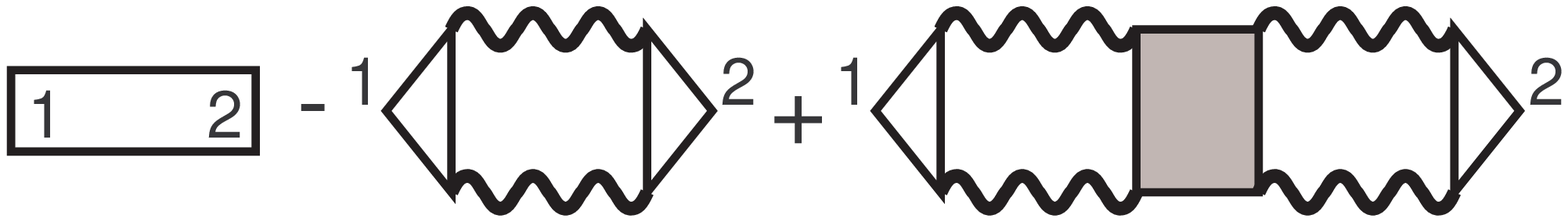}},
\end{equation}
where the full four-vertex is a solution of the Bethe--Salpeter type
equation
\begin{equation}\label{L-BS}
 \raisebox{-10pt}{\includegraphics[scale=0.3]{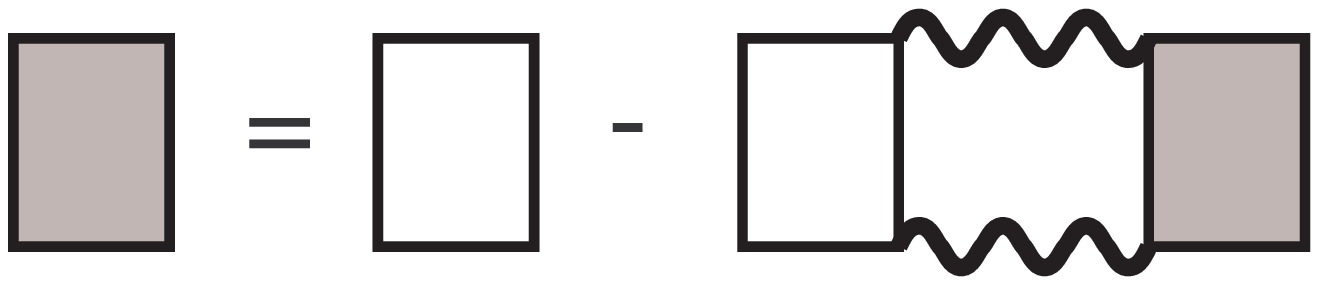}}.
\end{equation}
Here the thick wavy lines represent the renormalized hopping
$\Tilde{\mathbf{t}}_{\bm{k}}(i\omega_{\nu})$ [Eq. (\ref{tchains})], and
\begin{align}\label{vertdef}
  \raisebox{-4pt}{\includegraphics[scale=0.3]{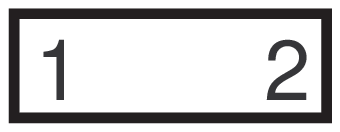}} & =-\frac{d^2\Omega_0}{d\mu_1d\mu_2}
  -\frac1N\frac{\partial^2\Tilde{\Phi}_{\text{lat}}}{\partial\mu_1\partial\mu_2}
  =\frac{\partial n_1}{\partial\mu_2}=\frac{\partial n_2}{\partial\mu_1},
  \nonumber\\
  \raisebox{-11pt}{\includegraphics[scale=0.3]{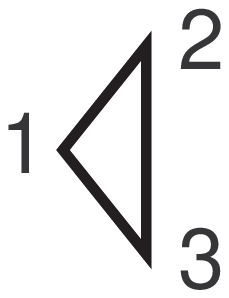}} & =
  \beta\frac{\partial}{\partial\mu_1}
  \frac{\delta\Tilde{\Phi}_{\text{lat}}}{\delta\Tilde{t}_{32}}
  =\frac{\partial\Xi_{23}}{\partial\mu_1}
  =\frac{\delta n_1}{\delta\Tilde{t}_{32}},\\
  \raisebox{-11pt}{\includegraphics[scale=0.3]{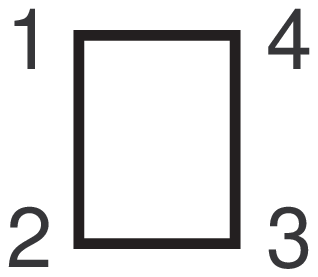}} & =
  -\beta\frac{\delta^2\Tilde{\Phi}_{\text{lat}}}{\delta\Tilde{t}_{21}\delta\Tilde{t}_{43}}
  =-\frac{\delta\Xi_{12}}{\delta\Tilde{t}_{43}}
  =-\frac{\delta\Xi_{34}}{\delta\Tilde{t}_{21}}
  \nonumber
\end{align}
are irreducible vertices that can not be divided into parts by cutting two
hopping lines, and it can be shown that in the $D\to\infty$ limit all
irreducible vertices become local.\cite{ShvaikaJPS}

\section{Falicov--Kimball model}

Now let us apply the developed above approach to the Falicov--Kimball
model with correlated hopping. In this case the single-site Hamiltonian
$H_i$ is written as
\begin{equation}\label{H_FK1}
  H_i=-\mu_f n_{if}-\mu_d n_{id}+Un_{id}n_{if}
\end{equation}
or
\begin{align}
  H_i&=\left[(U-\mu_d)n_{id} - \mu_f\right] P_i^+ -\mu_d n_{id}P_i^-
  \nonumber\\
  &=H_i^+P_i^++H_i^-P_i^-.
  \label{H_FK2}
\end{align}
Projection operators $P_i^+$ and $P_i^-$ commute with the total
Hamiltonian [Eq. (\ref{Htot})] and the statistical operator for the
effective single-impurity model [Eq. (\ref{rhoimp})] can be represented as
\begin{equation}\label{rhoproj}
  \Hat{\rho}_{\text{imp}}=P_o^+\Hat{\rho}_+ + P_o^-\Hat{\rho}_-,
\end{equation}
where
\begin{equation}\label{rhosub}
  \Hat{\rho}_{\alpha}=e^{-\beta H_o^{\alpha}}T\exp\left\{
  \!-\!\int_0^{\beta}\!d\tau\!\int_0^{\beta}\!d\tau'
  d_o^{\dag}(\tau)J^{\alpha\alpha}(\tau-\tau')d_o(\tau')
  \!\right\}
\end{equation}
are statistical operators for the \emph{noninteracting} $d$ electrons
placed in the different fields for the different subspaces $\alpha=\pm$.
As a result, the partition function for the impurity is a sum of the
partition functions for the subspaces:
\begin{equation}\label{Zimp}
  Z_{\text{imp}}=\Tr\Hat{\rho}_{\text{imp}}=e^{-\beta Q_+} + e^{-\beta Q_-},
\end{equation}
\begin{align}
  Q_+ = & -\mu_f - \frac1{\beta}\ln\left(1+e^{-\beta(U-\mu_d)}\right) \nonumber\\
        & - \frac1{\beta}\sum_{\nu}\ln\left(1-
          \frac{J^{++}(i\omega_{\nu})}{i\omega_{\nu}+\mu_d-U}\right),
  \label{Qdef}\\
  \nonumber
  Q_- = & - \frac1{\beta}\ln\left(1+e^{\beta\mu_d}\right)
         - \frac1{\beta}\sum_{\nu}\ln\left(1-
          \frac{J^{--}(i\omega_{\nu})}{i\omega_{\nu}+\mu_d}\right),
\end{align}
that gives, for the grand canonical potential of the impurity, the
following expression
\begin{equation}\label{OmegaFKimp}
  \Omega_{\text{imp}}=
  -\frac1{\beta}\ln\left(e^{-\beta Q_+} + e^{-\beta Q_-}\right).
\end{equation}

From Eq. (\ref{rhoproj}) it also follows that components of the Green's
function for the impurity are equal
\begin{align}
  G_{\text{imp}}^{++}(\omega) & =
  \frac{\langle P^+\rangle}{\omega+\mu_d - U - J^{++}(\omega)},
  \nonumber\\
  G_{\text{imp}}^{--}(\omega) & =
  \frac{\langle P^-\rangle}{\omega+\mu_d - J^{--}(\omega)},
  \label{GFimp}\\
  G_{\text{imp}}^{+-}(\omega) & = G_{\text{imp}}^{-+}(\omega) = 0.
  \nonumber
\end{align}
Here
\begin{equation}\label{Pplus}
  n_f=\langle P^+\rangle=
  \frac{e^{-\beta Q_+}}{e^{-\beta Q_+} + e^{-\beta Q_-}}
\end{equation}
is the $f$-particle concentration. In addition, for the $d$-particle
concentration we have
\begin{equation}\label{ndavg}
  n_d=\langle d^{\dag}d\rangle=\frac1{\beta}\sum_{\nu}
  \left[G_{\text{imp}}^{++}(i\omega_{\nu})
  +G_{\text{imp}}^{--}(i\omega_{\nu})\right].
\end{equation}

The coherent potential $\mathbf{J}(\omega)$ is a solution of Eq.
(\ref{DMFTeqJ}), and finally for the Green's function of the lattice we
obtain
\begin{widetext}
\begin{equation}\label{Glatfin}
  \mathbf{G}_{\bm{k}}(\omega)=
  \frac1{\mathcal{D}_{\bm{k}}(\omega)}
  \begin{Vmatrix}
  \left[\omega + \mu_d - J^{--}(\omega)\langle P^+\rangle
  -t_{\bm{k}}^{--}\langle P^-\rangle\right]\langle P^+\rangle
  &
  -\left[J^{+-}(\omega) - t_{\bm{k}}^{+-}\right]\langle P^+\rangle\langle P^-\rangle
  \\
  -\left[J^{-+}(\omega) - t_{\bm{k}}^{-+}\right]\langle P^+\rangle\langle P^-\rangle
  &
  \left[\omega + \mu_d - U - J^{++}(\omega)\langle P^-\rangle
  -t_{\bm{k}}^{++}\langle P^+\rangle\right]\langle P^-\rangle
  \end{Vmatrix},
\end{equation}
\begin{equation}\label{det}
  \begin{split}
  \mathcal{D}_{\bm{k}}(\omega)=&
  \left[\omega+\mu_d-J^{--}(\omega)\langle P^+\rangle
  -t_{\bm{k}}^{--}\langle P^-\rangle\right]
  \left[\omega+\mu_d-U-J^{++}(\omega)\langle P^-\rangle
  -t_{\bm{k}}^{++}\langle P^+\rangle\right]\\
  &-\left[J^{+-}(\omega) -t_{\bm{k}}^{+-}\right]
  \left[J^{-+}(\omega) -t_{\bm{k}}^{-+}\right]\langle P^+\rangle\langle
  P^-\rangle.
  \end{split}
\end{equation}
The inverse matrix of the irreducible part is equal,
\begin{equation}
  \mathbf{\Xi}^{-1}(\omega)=
  \begin{Vmatrix}
  \frac1{\langle P^+\rangle}\left[\omega+\mu_d-U-J^{++}(\omega)\langle P^-\rangle\right]
  & J^{+-}(\omega)\\
  J^{-+}(\omega) &
  \frac1{\langle P^-\rangle}\left[\omega+\mu_d-J^{--}(\omega)\langle P^+\rangle\right]
  \end{Vmatrix}
\end{equation}
\end{widetext}
and one can see that in the $|\omega|\to\infty$ limit its diagonal
components diverge faster than $\omega$ that does not allow to introduce a
matrix of the local (CPA) self-energies with correct analytic properties
by Eq. (\ref{locSigm}). As a rule, the behavior of the CPA self-energy at
the $\omega\to\pm\infty$ limit is not analyzed considering analytic
properties of the BEB CPA.\cite{BEB2,BEB3} On the other hand, the
components of the irreducible part matrix at $|\omega|\to\infty$ behave as
\begin{gather}
  \Xi^{++}(\omega)\to\frac{\langle P^{+}\rangle}{\omega}; \qquad
  \Xi^{--}(\omega)\to\frac{\langle P^{-}\rangle}{\omega},
  \nonumber\\
  \Xi^{+-}(\omega),\Xi^{-+}(\omega)\to \mathcal{O}\left(\frac{1}{\omega^3}\right),
  \label{Xilim}
\end{gather}
that is in agreement with Eq. (\ref{omginf}).

In the absence of $f$ particles $\langle P^+\rangle=0$ and $\langle
P^-\rangle=1$, and we obtain solution for the free electrons,
\begin{equation}\label{freeminus}
  G_{\bm{k}}(\omega)=
  G_{\bm{k}}^{--}(\omega)=\frac1{\omega+\mu_d-t_{\bm{k}}^{--}},
\end{equation}
whereas for the opposite case, when all sites are occupied by
$f$-particles $\langle P^+\rangle=1$ and $\langle P^-\rangle=0$,
\begin{equation}\label{freeplus}
  G_{\bm{k}}(\omega)=
  G_{\bm{k}}^{++}(\omega)=\frac1{\omega+\mu_d-U-t_{\bm{k}}^{++}}.
\end{equation}

In general the total Green's function [Eq. (\ref{GFtot})] for a
Falicov--Kimball model with correlated hopping can be written in the Dyson
representation as
\begin{equation}\label{GFtotD}
  G_{\bm{k}}(\omega)=\frac1{\omega+\mu_d-\Sigma_{\bm{k}}(\omega)-\Bar{t}_{\bm{k}}},
\end{equation}
where
\begin{equation}\label{teff}
  \Bar{t}_{\bm{k}}=t_{\bm{k}}^{++}\langle P^+\rangle^2
  +t_{\bm{k}}^{--}\langle P^-\rangle^2+
  \left(t_{\bm{k}}^{+-}+t_{\bm{k}}^{-+}\right)\langle P^+\rangle\langle P^-\rangle
\end{equation}
is the Hartree (mean-field) renormalized hopping already introduced by
Schiller,\cite{Schiller} and
\begin{align}\label{Sigmatot}
  \Sigma_{\bm{k}}(\omega) =& U\langle P^+\rangle
  +J_3(\omega) \langle P^+\rangle\langle P^-\rangle\\
  & +\frac{U_{1\bm{k}}(\omega)U_{2\bm{k}}(\omega) \langle P^+\rangle\langle P^-\rangle}
  {\omega+\mu_d - U\langle P^-\rangle-\Bar{J}(\omega)
  -t_{3\bm{k}}\langle P^+\rangle\langle P^-\rangle}
  \nonumber
\end{align}
is a momentum-dependent (nonlocal) self-energy. Here
\begin{align}
  U_{1\bm{k}}(\omega)=& U
  +\left[t_{\bm{k}}^{++}-t_{\bm{k}}^{-+}-J^{--}(\omega)+J^{-+}(\omega)\right]
   \langle P^+\rangle
   \nonumber\\
  &+\left[t_{\bm{k}}^{+-}-t_{\bm{k}}^{--}-J^{+-}(\omega)+J^{++}(\omega)\right]
   \langle P^-\rangle
   \label{Ueff}\\
  U_{2\bm{k}}(\omega)=& U
  +\left[t_{\bm{k}}^{++}-t_{\bm{k}}^{+-}-J^{--}(\omega)+J^{+-}(\omega)\right]
   \langle P^+\rangle
   \nonumber\\
   \nonumber
  &+\left[t_{\bm{k}}^{-+}-t_{\bm{k}}^{--}-J^{-+}(\omega)+J^{++}(\omega)\right]
   \langle P^-\rangle
\end{align}
describe renormalization of the interaction $U$, and
\begin{align}\label{tJ}
  \Bar{J}(\omega)=&J^{++}(\omega)\langle P^-\rangle^2
  +J^{--}(\omega)\langle P^+\rangle^2
  \\
  &+\left(J^{+-}(\omega)+J^{-+}(\omega)\right)\langle P^+\rangle\langle
  P^-\rangle,
  \nonumber\\
  J_3(\omega)=&J^{++}(\omega)+J^{--}(\omega)-J^{+-}(\omega)-J^{-+}(\omega),
  \\
  t_{3\bm{k}}=&t_{\bm{k}}^{++}+t_{\bm{k}}^{--}-
  t_{\bm{k}}^{+-}-t_{\bm{k}}^{-+}.
\end{align}

In his paper Schiller\cite{Schiller} considered the case of
\begin{equation}\label{t30}
  t_{3\bm{k}}=0,
\end{equation}
and from [Eq. (\ref{Sigmatot})] we obtain the same result for the
self-energy,
\begin{equation}\label{Sigma30}
  \Sigma_{\bm{k}}(\omega)=\Sigma_0(\omega)+
  \Sigma_1(\omega)t_{\bm{k}}+\Sigma_2(\omega)t_{\bm{k}}^2,
\end{equation}
that for the nearest-neighbor hopping $t_{\bm{k}}$ contains only local,
nearest-neighbor and next-nearest-neighbor contributions. The connection
with Schiller's approach for this case is given in the Appendix. But in a
general case of $t_{3\bm{k}}\ne0$, when Schiller's approach is not
applied, the self-energy is extended over the whole lattice, see
Ref.~\onlinecite{BEB3} for a comparison with the BEB CPA.

\begin{figure}[!htb]
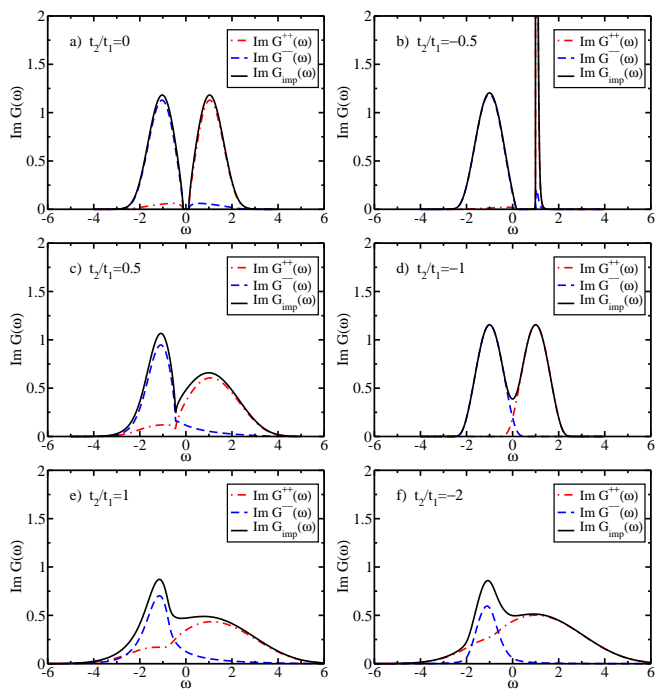

\centering
\includegraphics[width=0.49\columnwidth,clip]{t2_0}
\includegraphics[width=0.49\columnwidth,clip]{t2__05}
\\
\includegraphics[width=0.49\columnwidth,clip]{t2_05}
\includegraphics[width=0.49\columnwidth,clip]{t2__1}
\\
\includegraphics[width=0.49\columnwidth,clip]{t2_1}
\includegraphics[width=0.49\columnwidth,clip]{t2__2}
\caption{(Color online) The spectral function (imaginary part of the Green's function)
and its components for different relations $t_2/t_1$ ($t_3=0$) for the
$D=\infty$ hypercubic lattice with nearest-neighbor hopping ($W=1$, $U=2$,
$T=0.01$) at half-filling $\mu_d=\mu_f$, $n_f+n_d=1$.} \label{fig1}
\end{figure}

\begin{figure}[!htb]
\bigskip\centering
\includegraphics[width=0.9\columnwidth,clip]{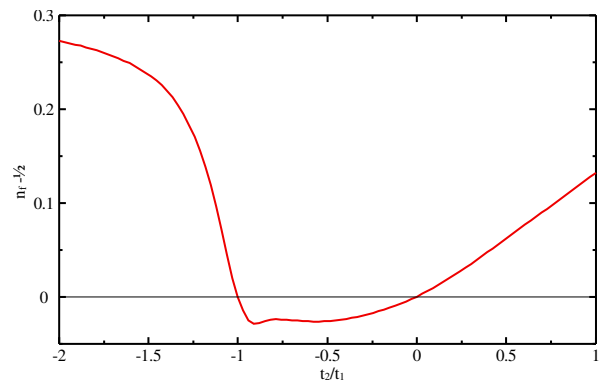}
\caption{(Color online) Deviation from the half-filling of the $f$-state occupation as a
function of $t_2/t_1$. Parameter values are the same as in
Fig.~\ref{fig1}.}\label{fig2}
\end{figure}

\begin{figure}[!htb]
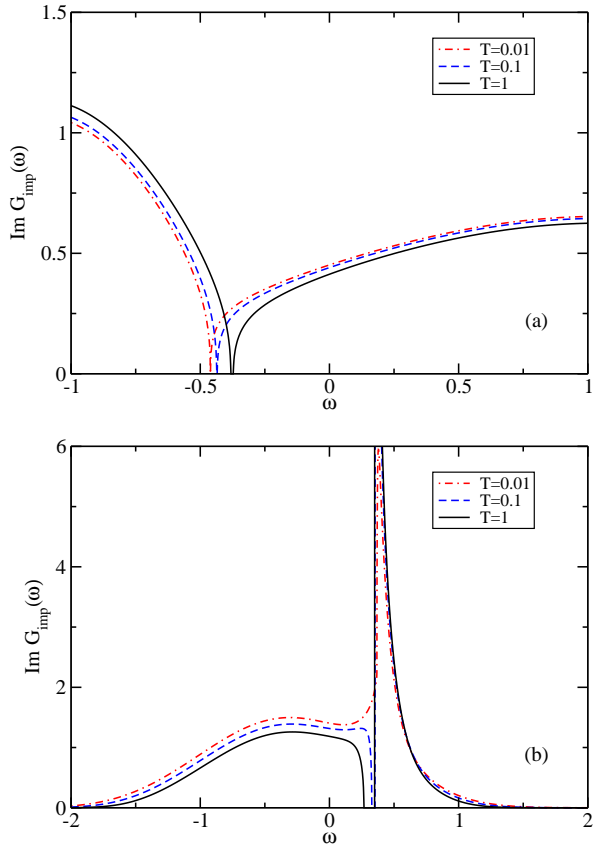

\centering
\includegraphics[width=0.9\columnwidth,clip]{U_21}\\ [1em]
\includegraphics[width=0.9\columnwidth,clip]{U_07}
\caption{(Color online) Temperature development of the gap (Mott transition) for (a)
$t_2/t_1=0.5$, $U=2.1$ and (b) $t_2/t_1=-0.5$, $U=0.7$.}
\label{fig3}
\end{figure}

In Fig.~\ref{fig1}
we present the spectral function (imaginary part of the Green's function)
and its components for different relations $t_2/t_1$ ($t_3=0$) for the
Gaussian density of states ($D=\infty$ hypercubic lattice with
nearest-neighbor hopping),
\begin{equation}\label{Gauss}
  \rho(t)=\frac1{W\sqrt{\pi}}e^{-\frac{t^2}{W^2}},
\end{equation}
($W=1$) at half-filling $\mu_d=\mu_f$, $n_f+n_d=1$. One can see that these
results are in good agreement with the one obtained by
Schiller\cite{Schiller} for the $D=2$ square lattice. At points
$t_2/t_1=0$ and $t_2/t_1=-1$ we have a symmetric band picture, whereas for
$t_2/t_1>0$ and $t_2/t_1<-1$ parameter values of the correlated hopping
the upper band become wider and for $-1<t_2/t_1<0$ values the lower band
does so. Such a behavior can be explained in the following way. The
symmetry of the band picture is determined by the relation
$|t^{++}|/|t^{--}|$ and by the deviation of the $f$-state occupation from
half-filling $n_f-\frac12$. For the symmetric points we have
$t^{++}=t^{--}$ for the $t_2/t_1=0$ case and $t^{++}=-t^{--}$ for the
$t_2/t_1=-1$ case. In both cases $n_f=\frac12$ for any temperature value,
but out of these two points there is a deviation from the half-filling of
the $f$-state occupation (see Fig.~\ref{fig2}),
and this deviation achieves its maximum at low temperatures. On the other
hand, at high temperature $n_f\to\frac12$ for $T\to\infty$. The gap in the
spectrum is at a maximum at $n_f\approx\frac12$, and with the decrease of
the temperature such a deviation from the half-filling of the $f$ states
can lead to a shutdown of the gap and a Mott transition (see
Fig.~\ref{fig3}).

Now let us consider some limiting cases. In the absence of correlated
hopping ($t_2=t_3=0$),
\begin{equation}\label{ch0t}
  t_{\bm{k}}^{++}=t_{\bm{k}}^{--}=t_{\bm{k}}^{+-}=t_{\bm{k}}^{-+}=t_{\bm{k}}
\end{equation}
and
\begin{equation}\label{ch0J}
  J^{++}(\omega)=J^{--}(\omega)=J^{+-}(\omega)=J^{-+}(\omega)=J(\omega),
\end{equation}
that follows from definition (\ref{CohPot}), and we obtain the known
result for the Falicov--Kimball model,
\begin{gather}\label{ch0}
  \Bar{t}_{\bm{k}}=t_{\bm{k}},\\
  \Sigma(\omega) = U\langle P^+\rangle
   +\frac{U^2\langle P^+\rangle\langle P^-\rangle}
  {\omega+\mu_d - U\langle P^-\rangle-J(\omega)}
\end{gather}
with local self-energy.

Another symmetric point $t_2/t_1=-1$ corresponds to the case of diagonal
hopping matrix ($t_{\bm{k}}^{+-}=t_{\bm{k}}^{-+}=0$) when hopping is
allowed only between sites with the same occupancy. In this case we have a
binary alloy with atoms of two types distributed stochastically over the
lattice, and hopping is restricted between atoms of different types. Now
atoms of one subsystem can be treated as impurities for the other
subsystem, and it was supposed in Ref.~\onlinecite{BEB} that this case
corresponds to two independent bands (independent band limit). In this
limit the particular case of $t_3=0$ was considered for the
one-dimensional Hubbard model when exact results can be
obtained.\cite{Strack} In Ref.~\onlinecite{Bulka} it was shown for the
Hubbard model, by the slave-boson mean-field approach, that in this case a
direct transition from the superconducting state to the Mott insulator
takes place.

A coherent potential matrix in this case is also diagonal
[$J^{+-}(\omega)=J^{-+}(\omega)=0$] and for the Falicov--Kimball model and the semielliptic density of states [Eq.
(\ref{Bethe})] we can obtain analytic expressions
\begin{align}
  J^{++}(\omega) &= \frac{a_{++}^2W^2}{4}G^{++}(\omega)
  \label{Jpp}\\ \nonumber
  &=\frac12\left[\omega+\mu_d-U\pm
  i\sqrt{W_+^2-(\omega+\mu_d-U)^2}\right],\\
  \label{Jmm}
  J^{--}(\omega) &= \frac{a_{--}^2W^2}{4}G^{--}(\omega)\\
  &=\frac12\left[\omega+\mu_d\pm
  i\sqrt{W_-^2-(\omega+\mu_d)^2}\right],
  \nonumber
\end{align}
\[
  W_{\pm}^2=a_{\pm\pm}^2W^2\langle P^{\pm}\rangle.
\]
The Green's function for the lattice contains two contributions
\begin{align}\label{gggg}
  G_{\bm{k}}^{++} & = \frac{\langle P^+\rangle}{\omega+\mu_d-U-
  J^{++}(\omega)\langle P^-\rangle-t_{\bm{k}}\langle P^+\rangle},\\
  G_{\bm{k}}^{--} & =\frac{\langle P^-\rangle}{\omega+\mu_d-
  J^{--}(\omega)\langle P^+\rangle-t_{\bm{k}}\langle P^-\rangle}
\end{align}
that describe two independent bands originating from two disordered
subsystems characterized by their own coherent potentials. The spectral
weight function contains two bands
\begin{align}\label{swf}
  \rho(\omega)=&\frac1{\pi}\Img G_{\text{imp}}(\omega-\mu-i0^+)
  \nonumber\\
  =& \langle P^+\rangle \frac2{\pi W_+^2}\sqrt{W_+^2-(\omega-U)^2}
  \nonumber\\
  &+\langle P^-\rangle \frac2{\pi W_-^2}\sqrt{W_-^2-\omega^2}
\end{align}
separated by the gap
\begin{equation}\label{gap}
  \Delta=U-|W_+|-|W_-|,
\end{equation}
that, in general, is temperature dependent and disappears at some critical
value
\begin{equation}\label{Uc}
  U_c=|W_+|+|W_-|=W\left(|a_{++}|\sqrt{\langle P^+\rangle}+
  |a_{--}|\sqrt{\langle P^-\rangle}\right)
\end{equation}
(the Mott transition point) that achieves its maximum
\begin{equation}
  U_c^{\text{max}}=W\sqrt{a_{++}^2+a_{--}^2}
\end{equation}
at
\begin{equation}
  n_f=\frac{a_{++}^2}{a_{++}^2+a_{--}^2}.
\end{equation}
At half-filling ($\mu_f=\mu_d$, $n_f+n_d=1$) and for the case $t_3=0$,
when $a_{--}=1$, $a_{++}=-1$, and $n_f=n_d=\frac12$, the gap does not
depend on temperature and disappears at $U_c=W\sqrt{2}$.

The same features can be observed in Fig.~\ref{fig1} [case (d) $t_2/t_1=-1$] for the Gaussian density of states too, when different subspaces contribute to different independent subbands. In this case there are no well-defined band edges and there are no exact gap for any value of $U$.

The next simplification of the model corresponds to a system with broken
bonds when we have hopping only between the sites not occupied by $f$
particles
\begin{gather}\label{bb}
  t_{\bm{k}}^{++}=t_{\bm{k}}^{+-}=t_{\bm{k}}^{-+}=0,\\
  J^{++}(\omega)=J^{+-}(\omega)=J^{-+}(\omega)=0.
\end{gather}
In other words, lattice sites are occupied stochastically by atoms of two
types (``conducting'' and ``isolating''), and electron hopping is allowed
only between atoms of the first type and the influence of other subsystem
is considered as disorder (the case of the ``diluted'' conductor). Now the
total Green's function [Eq. (\ref{GFtotD})] contains the renormalized
hopping
\[
\Bar{t}_{\bm{k}}=t_{\bm{k}}^{--}\langle P^-\rangle^2,
\]
and some momentum-dependent (nonlocal) self-energy
\begin{align}\label{bbSigma}
  &\Sigma_{\bm{k}}(\omega)=  U\langle P^+\rangle +
  J^{--}(\omega)\langle P^+\rangle\langle P^-\rangle \\
  & +\frac{\left[U-J^{--}(\omega)\langle P^+\rangle-
  t_{\bm{k}}^{--}\langle P^-\rangle\right]^2\langle P^+\rangle\langle P^-\rangle}
  {\omega+\mu_d-U\langle P^-\rangle-J^{--}(\omega)\langle P^+\rangle^2
  -t_{\bm{k}}^{--}\langle P^+\rangle\langle P^-\rangle}.
  \nonumber
\end{align}
But, in fact, these quantities have no any physical sense. Indeed, in this
case the total Green's function contains two contributions
\begin{align}\label{bbGF}
  G_{\bm{k}}(\omega) & =G_{\bm{k}}^{++}(\omega)+G_{\bm{k}}^{--}(\omega),\\
  G_{\bm{k}}^{++}(\omega) & = \frac{\langle P^+\rangle}
  {\omega+\mu_d-U}, \\
  G_{\bm{k}}^{--}(\omega) & = \frac{\langle P^-\rangle}
  {\omega+\mu_d-J^{--}(\omega)\langle P^+\rangle-t_{\bm{k}}^{--}\langle P^-\rangle},
\end{align}
where $G_{\bm{k}}^{++}(\omega)$ originates from the localized levels of
the isolating atoms and $G_{\bm{k}}^{--}(\omega)$ describes the
one-electron excitations in the disordered conducting system characterized
by the coherent potential $J^{--}(\omega)$. For the semielliptic density
of states [Eq. (\ref{Bethe})], we have
\begin{equation}\label{bbJse}
  J^{--}(\omega)=\frac12\left[\omega+\mu_d\pm i\sqrt{W^2\langle
  P^-\rangle-(\omega+\mu_d)^2}\right],
\end{equation}
and the one-particle spectral weight function
\begin{align}\label{bbdos}
  \rho(\omega)&=\frac1{\pi}\Img G_{\text{imp}}(\omega-\mu-i0^+)\\
  &=\langle P^+\rangle\delta(\omega-U)+
  \langle P^-\rangle\frac2{\pi W_-^2}\sqrt{W_-^2-\omega^2}
  \nonumber
\end{align}
contains the $\delta$-peek from the localized levels and conducting band
of the half-width $W_-=W\sqrt{\langle P^-\rangle}$ and their statistical
weights are equal to the concentrations of the isolating $\langle
P^+\rangle=n_f$ and conducting $\langle P^-\rangle=1-n_f$ atoms,
respectively. A critical value of $U$ when a ``gap'' develops is equal,
\begin{equation}\label{gap3}
  U_c=W_-=W\sqrt{\langle P^-\rangle},
\end{equation}
and is a function of the temperature. This limiting case of the
Falicov--Kimball model with correlated hopping shows that the explanation
of the model behavior in terms of the CPA approach is more appropriate
than in the Fermi-liquid language.

\section{Summary}

In this paper we presented a general approach to the description of
correlated hopping in dynamical mean-field theory, whose application is
not limited by the case of the Falicov--Kimball model and can be applied
to other models, e.g., Hubbard, $t$-$J$, etc. It is based on the Larkin
equation (an expansion over electron hopping around the atomic limit) that
considers all hopping terms, including correlated hopping, in the same
manner. Another starting point is the local character of the irreducible
part (irreducible cumulant) of the Green's functions, constructed by the
projected (Hubbard) operators, in the $D\to\infty$ limit, that is a more
general statement than the one about the local character of the
self-energy, which in the case of correlated hopping is nonlocal. Such an
approach keeps the dynamical mean-field theory local ideology, and allows
one to calculate the thermodynamical functions. To do this a grand
canonical potential functional of the Baym--Kadanoff-type, that allows one
to build a $\Phi$-derivatible theory without introducing the self-energy,
was proposed.

As an example, a Falicov--Kimball model with correlated hopping, and its
limiting case of a model with broken bonds (``diluted" conductor), were
considered when exact analytic results can be obtained. In particular, the
temperature-driven Mott transition in the Falicov--Kimball model with
correlated hopping was investigated. Such a strong coupling approach for
the Falicov--Kimball model with correlated hopping corresponds to the BEB
CPA approach for a binary alloy with off-diagonal disorder, and it follows
that the BEB CPA becomes exact in the $D=\infty$ limit. In addition,
analytic properties of the local self-energies in the BEB theory were
considered.

\begin{acknowledgments}
This work was supported by the Science and Technology Center in Ukraine
under grant No.~1673. We are grateful to I.V.~Stasyuk and J.~Freericks for
very useful and stimulating discussions. We thank O. Farenyuk for bringing the error in numerical results to our attention.
\end{acknowledgments}

\appendix*

\section{Equivalence to Schiller's approach}

For the case $t_{3\bm{k}}=0$ we can find a connection between components
of $\Sigma^S$ and $\mathbf{J}$ (here and below we use $J^{+-}=J^{-+}$ and
$\Sigma_{\psi d}^S=\Sigma_{d\psi}^S$) by comparison of Eqs.
(\ref{Sigma30}) and (S.8) [here and below (S.8) indicates Eq. (8) in the
paper by Schiller\cite{Schiller}]:
\begin{align}
  J^{++}&=\omega-U+\left(\Sigma_{dd}^S-U-
  \frac{(\Sigma_{\psi d}^S-t_2)^2}{\Sigma_{\psi\psi}^S}\right)
  \frac{n_f}{1-n_f},
  \nonumber\\
  J^{--}&=\omega+\left(\Sigma_{dd}^S-U-
  \frac{\left(\Sigma_{\psi d}^S\right)^2}{\Sigma_{\psi\psi}^S}\right)
  \frac{1-n_f}{n_f},
  \label{JvsS}\\
  J^{+-}&=\omega-\Sigma_{dd}^S+
  \frac{\Sigma_{\psi d}^S}{\Sigma_{\psi\psi}^S}(\Sigma_{\psi d}^S-t_2).
  \nonumber
\end{align}
In addition, according to their definition the components of the Green's
function (S.9) are not independent and satisfy the following relation:
\begin{equation}\label{identity}
  (\omega-\Sigma_{dd}^S)G_{dd}-(t_1+2\Sigma_{\psi d}^S)G_{\psi d}
  -\Sigma_{\psi\psi}^S G_{\psi\psi}=1.
\end{equation}
On the other hand, from Eqs. (\ref{DMFTeq1}), (\ref{DMFTeq2}),
(\ref{Glatfin}), and (\ref{JvsS}), we obtain a connection between the
components of the Green's functions
\begin{align}
  G^{++}&=\frac{\Sigma_{\psi\psi}^S}{t_2^2}\left[
  \left(\omega-\Sigma_{dd}^S+\frac{\left(\Sigma_{\psi d}^S\right)^2}{\Sigma_{\psi\psi}^S}\right)G_{dd}
  -t_1G_{\psi d}\right]
  \nonumber\\
  &=\frac{1-n_f}{U-\Sigma_{dd}^S+\frac{(\Sigma_{\psi d}^S-t_2)^2}{\Sigma_{\psi\psi}^S}},
  \nonumber\\
  G^{--}&=\frac{\Sigma_{\psi\psi}^S}{t_2^2}\left[
  \left(\omega-\Sigma_{dd}^S+\frac{(\Sigma_{\psi d}^S-t_2)^2}{\Sigma_{\psi\psi}^S}\right)G_{dd}
  \right.
  \nonumber\\
  &\left.\vphantom{\frac{(\Sigma_{\psi d}^S-t_2)^2}{\Sigma_{\psi\psi}^S}}
  -(t_1+2t_2)G_{\psi d}\right]
  =\frac{n_f}{\frac{\left(\Sigma_{\psi d}^S\right)^2}{\Sigma_{\psi\psi}^S}-\Sigma_{dd}^S},
  \label{GvsG}\\
  G^{+-}&=\frac{\Sigma_{\psi\psi}^S}{t_2^2}\left[(t_1+t_2)G_{\psi d}
  \vphantom{\frac{\Sigma_{\psi d}^S}{\Sigma_{\psi\psi}^S}}\right.
  \nonumber\\
  &-\left.
  \left(\omega-\Sigma_{dd}^S+
  \frac{\Sigma_{\psi d}^S}{\Sigma_{\psi\psi}^S}(\Sigma_{\psi d}^S-t_2)\right)
  G_{dd}\right]=0.
  \nonumber
\end{align}
A solution of this system of equations with the use of Eq.
(\ref{identity}) gives
\begin{align}
  G_{dd}&=G^{++}+G^{--},
  \nonumber\\
  G_{\psi d}&= -\frac{\Sigma_{\psi d}^S-t_2}{\Sigma_{\psi\psi}^S}G^{++}
  -\frac{\Sigma_{\psi d}^S}{\Sigma_{\psi\psi}^S}G^{--},
  \label{GvsGsol}\\
  G_{\psi\psi}&= \frac{\Sigma_{dd}^S-U}{\Sigma_{\psi\psi}^S}G^{++}
  +\frac{\Sigma_{dd}^S}{\Sigma_{\psi\psi}^S}G^{--}
  \nonumber
\end{align}
that is equivalent to the solution of Eqs. (S.5) and (S.10) with respect
to $G$:
\begin{equation}
  \Hat{G}=-n_f\left(\Hat{\Sigma}^S\right)^{-1}-(1-n_f)\left[\Hat{\Sigma}^S-
  \begin{pmatrix}U&t_2\\t_2&0\end{pmatrix}\right]^{-1}.
\end{equation}

For the $f$-particle occupation number $n_f=1/(e^{\beta\epsilon}+1)$,
where $\epsilon$ is defined by Eq. (S.11), we have
\begin{align}
  \epsilon&-E_f+\mu-\frac U2
  \nonumber\\
  &=-\frac1{\beta}\sum_{\nu}\ln\frac{n_f^2\det\left[
  \begin{pmatrix}U&t_2\\t_2&0\end{pmatrix}-\Hat\Sigma^S(i\omega_{\nu})\right]}
  {(1-n_f)^2\det\Hat\Sigma^S(i\omega_{\nu})}
  \nonumber\\
  &=-\frac1{\beta}\sum_{\nu}\ln\frac{i\omega_{\nu}+\mu-U-J^{++}(i\omega_{\nu})}
  {i\omega_{\nu}+\mu-J^{--}(i\omega_{\nu})}.
\end{align}
On the other hand, from Eqs. (\ref{Pplus}) and (\ref{Qdef}) we have
\begin{align}
  \epsilon&=Q_+-Q_-=-\mu_f-\frac1{\beta}\ln\frac{1+e^{-\beta(U-\mu_d)}}{1+e^{\beta\mu_d}}
  \\
  \nonumber
  &-\frac1{\beta}\sum_{\nu}\ln\frac{i\omega_{\nu}+\mu_d-U-J^{++}(i\omega_{\nu})}
  {i\omega_{\nu}+\mu_d-J^{--}(i\omega_{\nu})}
  \frac{i\omega_{\nu}+\mu_d}{i\omega_{\nu}+\mu_d-U}
\end{align}
that gives the same result if we use
\begin{equation}
  \frac1{\beta}\sum_{\nu}\ln\frac{i\omega_{\nu}+\mu_d-U}{i\omega_{\nu}+\mu_d}=
  \frac1{\beta}\ln\frac{1+e^{-\beta(U-\mu_d)}}{1+e^{\beta\mu_d}}.
\end{equation}

\end{document}